# Het 'right to be forgotten' en bijzondere persoonsgegevens: geen ruimte meer voor een belangenafweging?

# The 'right to be forgotten' and sensitive personal data: no room for balancing?

Dr. Frederik Zuiderveen Borgesius[1]

IViR, Universiteit van Amsterdam. f.j.zuiderveenborgesius [at] uva.nl

**Abstract (English) –** An attorney submitted a 'right to be forgotten' delisting request to Google, regarding a blog post about a criminal conviction of the attorney in another country. The Rotterdam District Court ruled that Google may no longer link to the blog post when people search for the attorney's name. The court granted the attorney's request because the blog post concerns a criminal conviction. Personal data regarding criminal convictions are, under Dutch law, special categories of data (sometimes called sensitive data). The reasoning of the court on special categories of data creates problems for freedom of expression. This paper, in Dutch, explores how these problems can be reduced. Google has appealed the decision; the judgment of the Court of Appeals is expected in March 2017.

**Abstract –** Een advocaat heeft een 'right to be forgotten'-verzoek gedaan bij Google, met betrekking tot een blogpost over een strafrechtelijke veroordeling van de advocaat in het buitenland. De Rechtbank Rotterdam heeft beslist dat Google niet meer naar de blogpost mag verwijzen als mensen zoeken op de naam van de advocaat. De rechtbank wees het verwijderingsverzoek toe omdat de blogpost een strafrechtelijke veroordeling betreft: een bijzonder persoonsgegeven. De redenering van de rechtbank over bijzondere persoonsgegevens leidt tot problemen voor de vrijheid van meningsuiting. Deze bijdrage verkent hoe die problemen verkleind kunnen worden.

## 1 Inleiding

Een man is in 2012 in het buitenland vanwege wapenbezit veroordeeld tot voorwaardelijke gevangenisstraf en een taakstraf. Een blogger uit dat land heeft toen over de zaak bericht, inclusief de naam en een foto van de man. Wanneer iemand de naam van de man in Google intypte, liet Google onder meer links zien naar de blogpost over de veroordeling. De man is nu advocaat (of hij in 2012 al advocaat was, blijkt niet uit het vonnis.)

Met een beroep op het Google Spain-arrest van het Hof van Justitie van de Europese Unie (HvJ EU) deed de advocaat een verwijderingsverzoek bij Google. Google weigerde de link te verwijderen. De advocaat stapte daarop naar het College Bescherming Persoonsgegevens, met het verzoek voor hem te bemiddelen.[2] Het College wees het verzoek tot bemiddeling af omdat de veroordeling vrij recent is, en berichtgeving daarover in het publiek belang is.[3] Vervolgens ging de advocaat naar de rechter; daar had hij meer succes. Dit voorjaar besliste de Rechtbank Rotterdam dat Google niet meer naar de blogpost mocht verwijzen als mensen in Google zochten op de naam van de advocaat, omdat het een bijzonder persoonsgegeven betreft, namelijk een gegeven over een strafrechtelijke veroordeling.[4]

---

[2] Zie artikel 47 Wbp. Het College Bescherming Persoonsgegevens heet inmiddels de Autoriteit Persoonsgegevens. Zie voor een overzicht van 'right to be forgotten' beslissingen door de Autoriteit Persoonsgegevens: 'Overzicht bemiddeling Autoriteit Persoonsgegevens bij verwijdering zoekresultaten Google', 25 mei 2016, https://autoriteitpersoonsgegevens.nl.
[3] Rechtbank Rotterdam, 29 maart 2016, ECLI:NL:RBROT:2016:2395, r.o. 2.7.

Hierna bespreek ik hoe de Rechtbank Rotterdam een afweging maakt tussen informatievrijheid en privacy (2). Die afweging is verdedigbaar. Vervolgens (3) behandel ik hoe het Google Spain-arrest een probleem heeft veroorzaakt met betrekking tot bijzondere persoonsgegevens. Daarna (4) bespreek ik hoe de Rotterdamse rechtbank de regels voor bijzondere gegevens toepast, en daarmee het probleem vergroot.[5] Er is kennelijk geen ruimte voor een belangenafweging als er bijzondere persoonsgegevens in het spel zijn. Bovendien lijkt de zoekmachine van Google nu grotendeels illegaal te zijn. Tot slot (5) verken ik hoe het probleem verkleind zou kunnen worden, bijvoorbeeld door de rechter in hoger beroep. Ik stip een aantal mogelijke oplossingen aan; aan elke oplossing kleven nadelen.

## 2    Rotterdamse vonnis: belangenafweging

Zaken waarbij privacy en vrijheid van meningsuiting afgewogen moeten worden zijn vaak moeilijk. Het recht op vrijheid van meningsuiting omvat volgens het Europees Verdrag voor de Rechten van de Mens, en volgens het Handvest van de Grondrechten van de Europese Unie, ook het recht om informatie te ontvangen: 'de vrijheid om inlichtingen of denkbeelden te ontvangen of te verstrekken'.[6] Kortheidshalve gebruik ik hier ook de term informatievrijheid in plaats van het recht op vrijheid van meningsuiting.[7]

---

[4] Rechtbank Rotterdam, 29 maart 2016, ECLI:NL:RBROT:2016:2395.
[5] De rechtbank Rotterdam beslist ook dat Google niet alleen Google.nl moet opschonen, maar ook Google.com. Een bespreking van dat aspect valt buiten het bestek van deze bijdrage. Zie daarover: C. Piltz, 'Google challenges fine by French Data Protection Authority: a "right to be forgotten" for the entire world?', 20 mei 2016, www.delegedata.de; Y. Fouad, 'Reikwijdte van het Europese dataprotectierecht na Google Spanje: wat is de territoriale werkingssfeer en wordt eenieder beschermd?', masterscriptie IViR/UvA 2015.
[6] Art. 11 Handvest; art. 10 lid 1 EVRM.
[7] De Rechtbank Rotterdam spreekt ook van een 'recht op informatie', Rechtbank Rotterdam, 29 maart 2016, ECLI:NL:RBROT:2016:2395, r.o. 4.8.

Sinds het Google Spain-arrest van het HvJ EU hebben betrokkenen, onder bepaalde omstandigheden, het recht Google een link niet te laten tonen in de zoekresultaten, als in Google wordt gezocht op de naam van de betrokkene.[8] Volgens het HvJ EU moet bij zo'n verwijderingsverzoek een afweging gemaakt worden tussen het recht op privacy en de bescherming van persoonsgegevens van de betrokkene, het belang van zoekmachinegebruikers om informatie te vinden, en het economische belang van de zoekmachine-aanbieder.[9] Mensen die een rol spelen in het openbare leven, zoals politici, maken minder kans op een succesvol verwijderingsverzoek.[10]

De auteur van de originele informatie, zoals een blogger of een journalist, heeft recht op vrijheid van meningsuiting. Als Google een link naar een publicatie verwijdert (voor zoekopdrachten op naam), is die publicatie moeilijker te vinden. Een publicatie moeilijker vindbaar maken kan gezien worden als het inperken van de vrijheid van meningsuiting van de originele auteur.[11]

In de zaak van de advocaat en het verboden wapenbezit zijn er argumenten die tegen verwijdering pleiten. Verboden wapenbezit is een strafbaar feit. In 2014 zei de Rechtbank Amsterdam in een zaak waarin een veroordeelde crimineel een verwijderingsverzoek bij Google had ingediend: 'Het plegen van een misdrijf heeft nu eenmaal tot gevolg dat men op zeer negatieve wijze in het nieuws kan komen en dit laat ook op het internet – mogelijk zelfs zeer langdurig – zijn sporen na.'[12] De rechtbank voegde daaraan toe dat 'de veroordeling voor een ernstig misdrijf zoals het onderhavige en de negatieve publiciteit als gevolg daarvan (…) in het algemeen blijvend relevante informatie over een persoon' zijn.[13] Van de rechter mocht Google het verwijderingsverzoek dus afwijzen. In hoger beroep besliste het gerechtshof in die zaak ook in het voordeel van Google. Het Hof zei: 'Het publiek heeft in het algemeen een groot belang om toegang te verkrijgen tot informatie omtrent ernstige delicten en dus ook omtrent de strafvervolging en veroordeling van [appellant]'.[14]

---

[8] HvJ EU 13 mei 2014, C-131/12, ECLI:EU:C:2014:317 (Google Spain). Zie uitgebreid: J. Ausloos en A. Kuczerawy, 'From Notice-and-Takedown to Notice-and-Delist: Implementing the Google Spain Ruling', 5 oktober 2015, *ICRI Research Paper 24* www.ssrn.com.
[9] HvJ EU 13 mei 2014, C-131/12, ECLI:EU:C:2014:317 (Google Spain), par. 81. De ondernemingsvrijheid van Google valt buiten het bestek van deze bijdrage.
[10] Zie HvJ EU 13 mei 2014, C-131/12, ECLI:EU:C:2014:317 (Google Spain). Par. 99.
[11] Zie: J. van Hoboken, *Search Engine Freedom. On the Implications of the Right to Freedom of Expression for the Legal Governance of Web Search Engines*, Alphen aan den Rijn: Kluwer Law International 2012, p. 350; S. Kulk & F.J. Zuiderveen Borgesius, 'De implicaties van het Google Spain-arrest voor de vrijheid van meningsuiting', *NTM/NJCM-Bull*. 2015, 40-1.
[12] Rechtbank Amsterdam, 18 September 2014, ECLI:NL:RBAMS:2014:6118, r.o. 4.11.
[13] Rechtbank Amsterdam, 18 September 2014, ECLI:NL:RBAMS:2014:6118, r.o. 4.11.

Maar in de zaak van de advocaat zijn er ook argumenten die vóór verwijdering pleiten. De Artikel 29-werkgroep, een samenwerkingsverband van Europese privacytoezichthouders, zegt dat een verwijderingsverzoek met betrekking tot een licht vergrijp van lang geleden in beginsel eerder toegekend moet worden dan een verzoek met betrekking tot een recente zware misdaad.[15] Het verboden wapenbezit van de advocaat was redelijk recent (2012), maar verboden wapenbezit is een lichter vergrijp dan een poging tot het huren van een moordenaar (waar de bovenstaande Amsterdamse zaak over ging).

Bovendien heeft de advocaat tijdens de Rotterdamse rechtszaak bewijs aangedragen dat in het land waar hij veroordeeld was, het strafblad inmiddels geschoond was ('conviction spent'), omdat mensen daar op grond van de wet na een veroordeling op een gegeven moment met een schone lei mogen beginnen.[16] Dat is voor de rechtbank een reden om het zoekresultaat in kwestie als niet meer relevant te zien.[17]

Verder vindt de rechtbank een advocaat niet per definitie een publiek figuur. Het gaat 'de rechtbank te ver om er vanuit te gaan dat iedere advocaat een zodanige maatschappelijke rol heeft dat het publiek steeds belang heeft om ervan kennis te kunnen nemen dat een advocaat strafrechtelijk is veroordeeld.'[18]

---

[14] Gerechtshof Amsterdam, 31 Maart 2015, ECLI:NL:GHAMS:2015:1123 (r.o. 3.6). Zie ook: Rechtbank Amsterdam, 24 december 2015, ECLI:NL:RBAMS:2015:9515 (over een journalist die 16 jaar geleden beschuldigd was van plagiaat).
[15] Artikel 29-werkgroep, 'Guidelines on the implementation of the Court of Justice of the European Union judgment on 'Google Spain and inc v. Agencia Española de Protección de Datos (AEPD) and Mario Costeja González' C-131/121', 26 november 2014, p. 20. 'As a rule, DPAs are more likely to consider the de-listing of search results relating to relatively minor offences that happened a long time ago, whilst being less likely to consider the de-listing of results relating to more serious ones that happened more recently'.
[16] Rechtbank Rotterdam, 29 maart 2016, ECLI:NL:RBROT:2016:2395, r.o. 4.11.6.
[17] Zie ook: Artikel 29-werkgroep, 'Guidelines on the implementation of the Court of Justice of the European Union judgment on 'Google Spain and inc v. Agencia Española de Protección de Datos (AEPD) and Mario Costeja González' C-131/121', 26 november 2014, p. 20.
[18] Rechtbank Amsterdam, 18 September 2014, ECLI:NL:RBAMS:2014:6118, r.o. 4.11.5.

Volgens de rechtbank is ook relevant dat de advocaat werkt op het gebied van zakelijke contracten. Het verboden wapenbezit houdt geen verband met zijn werk. Misschien zou de rechtbank anders beslist hebben als het om een strafrechtadvocaat ging. Alles in overweging nemend, concludeert de rechtbank dat de privacy van de advocaat in dit geval zwaarder weegt dan het publiek belang bij het kunnen vinden van de blogpost bij zoekopdrachten op de naam van de advocaat.

De afweging door de rechtbank is begrijpelijk. Maar, voor de rechtbank is deze afweging subsidiair. De rechtbank volgt primair een andere redenering, die ook uitpakt in het voordeel van de advocaat. Die primaire redenering van de rechtbank is gebaseerd op het wettelijk regime voor het verwerken van bijzondere persoonsgegevens. De redenering is, hoewel formeel juridisch te volgen, problematisch. Hierna bespreek ik eerst de achtergrond van het bijzondere-persoonsgegevens-probleem, en daarna het probleem na het Rotterdamse vonnis.

## 3    Google Spain-arrest: het bijzondere-persoonsgegevens-probleem

Sinds het Google Spain-arrest van het HvJ EU is er een probleem met bijzondere persoonsgegevens. Het HvJ EU heeft beslist dat Google een 'verantwoordelijke' is, kort gezegd omdat Google webpagina's met persoonsgegevens indexeert, en naar die pagina's verwijst met haar zoekmachine.[19] Volgens de Wet bescherming persoonsgegevens (Wbp), en volgens het EU Handvest, mogen verantwoordelijken alleen persoonsgegevens verwerken met toestemming van de betrokkene, of op basis van een andere wettelijke grondslag.[20]

Het HvJ EU zegt dat Google zich voor het verwerken van persoonsgegevens voor haar zoekmachine kan beroepen op de legitiem-belang-bepaling (artikel 8(f) Wbp).[21] Kort gezegd staat de legitiem-belang-bepaling verwerking toe als de belangen van de verantwoordelijke zwaarder wegen dan de belangen van de betrokkene. In de woorden van de Wbp: persoonsgegevens mogen verwerkt worden als 'de gegevensverwerking noodzakelijk is voor de behartiging van het gerechtvaardigde belang van de verantwoordelijke (…), tenzij het belang of de fundamentele rechten en vrijheden van de betrokkene, in het bijzonder het recht op bescherming van de persoonlijke levenssfeer, prevaleert.'[22] Omdat Google zich op de legitiem-belang-bepaling kan beroepen, mag Google persoonsgegevens verwerken voor haar zoekmachine, zonder toestemming te vragen aan alle betrokkenen.

---

[19] HvJ EU 13 mei 2014, C-131/12, ECLI:EU:C:2014:317 (Google Spain), par. 33
[20] Artikel 8 Wbp; artikel 8(2) van het Handvest.
[21] HvJ EU 13 mei 2014, C-131/12, ECLI:EU:C:2014:317 (Google Spain), par. 73.

Maar voor bijzondere persoonsgegevens geldt de legitiem-belang-bepaling niet. Bijzondere persoonsgegevens zijn onder meer gegevens over strafrechtelijke veroordelingen. De verwerking van bijzondere gegevens is in beginsel verboden, tenzij een wettelijke uitzondering van toepassing is als bedoeld in artikel 17 t/m 23 Wbp, óf als de betrokkene uitdrukkelijke toestemming verleent.[23] De hoofdregel uit de Wbp leest als volgt.

> De verwerking van persoonsgegevens betreffende iemands godsdienst of levensovertuiging, ras, politieke gezindheid, gezondheid, seksuele leven, alsmede persoonsgegevens betreffende het lidmaatschap van een vakvereniging is verboden behoudens het bepaalde in deze paragraaf [van de Wbp]. Hetzelfde geldt voor strafrechtelijke persoonsgegevens en persoonsgegevens over onrechtmatig of hinderlijk gedrag in verband met een opgelegd verbod naar aanleiding van dat gedrag.[24]

In veel gevallen bevatten websites bijzondere persoonsgegevens, en vaak verwijst Google naar zulke websites (overigens zonder onderscheid te maken tussen gewone en bijzondere gegevens). Bijvoorbeeld, als mensen identificeerbaar zijn op foto's en video's, betreffen die beelden persoonsgegevens. Vaak zal het ras (de huidskleur) van de betrokkene te zien zijn. In de media wordt vaak bericht over strafrechtelijke veroordelingen. En een ledenlijst op de website van een katholieke zangvereniging of een kerk is een lijst met bijzondere gegevens (over religie).

---

[22] Artikel 8(f) Wbp; artikel 7(f) Richtlijn bescherming persoonsgegevens.
[23] Artikel 16 Wbp. Bijzondere persoonsgegevens worden soms gevoelige gegevens genoemd, maar dat is niet helemaal juist. Alle persoonsgegevens kunnen meer of minder gevoelig zijn. Alleen een aantal categorieën persoonsgegevens zijn gedefinieerd als 'bijzondere gegevens' in artikel 16 Wbp. Zie Artikel 29-werkgroep, 'Advies 06/2014 over het begrip "gerechtvaardigd belang van de voor de gegevensverwerking verantwoordelijke" in artikel 7 van Richtlijn 95/46/EG' (WP 217), Brussel, 9 april 2014, p. 46-47.
[24] Artikel 16 Wbp; toevoeging auteur.

Google kan zich niet beroepen op een uitzondering in de Wbp voor het verwerken van bijzondere persoonsgegevens. Als Google betrokkenen om hun uitdrukkelijke toestemming zou vragen voor het verwerken van hun bijzondere persoonsgegevens, zou dat in theorie een verwerkingsgrondslag kunnen bieden. Maar dat zou ondoenlijk zijn. Google kan evenmin van te voren verifiëren of de uitzondering van toepassing is voor gevallen waarin mensen hun eigen bijzondere gegevens duidelijk openbaar hebben gemaakt.[25]

Kortom, met het Google Spain-arrest heeft het HvJ EU een probleem veroorzaakt.[26] Advocaat Generaal Jääskinen had al gewaarschuwd voor deze 'absurde' situatie.[27] We zouden ook kunnen zeggen dat het probleem niet is veroorzaakt door het HvJ EU, maar door een weeffout in de Richtlijn bescherming persoonsgegevens, waarop de Wbp is gebaseerd.[28] Het regime voor bijzondere gegevens komt immers uit de richtlijn. In ieder geval heeft het Google Spain-arrest het probleem op scherp gezet.

Sinds dat arrest hebben Google, privacytoezichthouders, en Nederlandse rechters het probleem met bijzondere gegevens opgelost door het te negeren. Zo heeft het Hof Amsterdam beslist dat een veroordeelde crimineel zich niet op het verwijderingsrecht kon beroepen. Kennelijk vond het hof dat Google de bijzondere persoonsgegevens

---

mocht verwerken.[29] De Artikel 29-werkgroep benoemde het bijzondere-persoons-gegevens-probleem ook niet in zijn opinie over het Google Spain-arrest.[30]

## 4  Rotterdamse vonnis: het bijzondere-persoonsgegevens-probleem

De advocaat in de zaak voor de Rechtbank Rotterdam zegt dat Google de gegevens over hem niet mag verwerken (en dus niet naar de blogpost mag verwijzen), simpelweg omdat het bijzondere persoonsgegevens betreft.[31] De rechtbank is het met dit argument eens:

> Nu het verzoek strafrechtelijke persoonsgegevens betreft, geldt in beginsel het verbod van verwerking van deze gegevens zoals is bepaald in artikel 16 Wbp. Voorts is niet gebleken van een van de uitzonderinggronden die limitatief en exclusief zijn opgesomd in artikel 22 Wbp op de toepassing van het verbod tot verwerking van de strafrechtelijke persoonsgegevens. <u>Hoewel de rechtbank zich bewust is van het vergaande praktische gevolg voor de verwerking van strafrechtelijke persoonsgegevens door de exploitant van een zoekmachine, acht de rechtbank de conclusie onvermijdelijk dat er in dit geval sprake is van een toepasselijk verbod op de verwerking van strafrechtelijke persoonsgegevens.</u> De rechtbank acht derhalve het verzoek om Google te bevelen de verwijzing naar de weblinks die voortkomen uit de zoekopdracht naar de naam van verzoeker uit de zoekresultaten te verwijderen dan wel af te schermen, reeds hierom toewijsbaar.[32]

---

[29] Gerechtshof Amsterdam, 31 Maart 2015, ECLI:NL:GHAMS:2015:1123.
[30] Artikel 29-werkgroep, 'Guidelines on the implementation of the Court of Justice of the European Union judgment on 'Google Spain and inc v. Agencia Española de Protección de Datos (AEPD) and Mario Costeja González' C-131/121', 26 november 2014.
[31] Rechtbank Rotterdam, 29 maart 2016, ECLI:NL:RBROT:2016:2395, r.o. 3.1.2.
[32] Rechtbank Rotterdam, 29 maart 2016, ECLI:NL:RBROT:2016:2395, r.o. 4.10.4. Nadruk toegevoegd.

Deze redenering leidt tot problemen. Ten eerste is de zoekmachine van Google nu grotendeels illegaal. Google verwijst immers naar veel webpagina's waarop bijzondere persoonsgegevens staan. De rechtbank lijkt zich te realiseren dat er een probleem ontstaat; de rechtbank is 'zich bewust (…) van het vergaande praktische gevolg voor de verwerking van strafrechtelijke persoonsgegevens door de exploitant van een zoekmachine.'[33] De conclusie dat Google's zoekmachine grotendeels illegaal is, is onwenselijk.[34] Toegegeven: zo lang de Wbp op dit punt niet gehandhaafd wordt tegen Google, is het probleem te overzien.

Als de redenering van de Rechtbank Rotterdam logisch wordt doorgezet, mag Google helemaal niet meer naar de blogpost met de bijzondere persoonsgegevens over de advocaat verwijzen. Persoonsgegevens verwerken zonder wettelijke grondslag is immers verboden. Dat verbod geldt ook als Google naar websites met bijzondere persoonsgegevens verwijst. Uit het Google Spain-arrest blijkt dat het indexeren van en verwijzen naar websites met persoonsgegevens gekwalificeerd moet worden als het verwerken van persoonsgegevens.[35]

Het tweede probleem is dat de Rechtbank Rotterdam geen ruimte ziet voor een belangenafweging: het is 'onvermijdelijk' dat het verbod geldt voor het verwerken van bijzondere gegevens.[36] De gedachtegang van de Rechtbank Rotterdam volgend, heeft iedereen over wie Google bijzondere persoonsgegevens verwerkt een absoluut verwijderingsrecht. Er is in zulke gevallen geen ruimte meer voor het meewegen van de informatievrijheid.

Stel nu dat een journalist naar waarheid op een website schrijft dat een politicus is veroordeeld wegens corruptie. Voor dit voorbeeld gaan we er van uit dat de publicatie een publiek belang dient. Volgens de gedachtegang van de Rechtbank Rotterdam heeft de politicus een absoluut verwijderingsrecht, en zou hij kunnen eisen dat Google de zoekresultaten op de naam van de politicus opschoont.

---

[33] Rechtbank Rotterdam, 29 maart 2016, ECLI:NL:RBROT:2016:2395, r.o. 4.10.4.
[34] Zie Opinie AG Jääskinen 25 juni 2013, ECLI:EU:C:2013:424 (Google Spain), par. 90.
[35] HvJ EU 13 mei 2014, C-131/12, ECLI:EU:C:2014:317 (Google Spain). Par. 41.
[36] Rechtbank Rotterdam, 29 maart 2016, ECLI:NL:RBROT:2016:2395, r.o. 4.10.4.

Maar de bescherming van privacy en persoonsgegevens zou niet absoluut moeten zijn. Zoals het Europese Hof voor de Rechten van de Mens opmerkt, hebben privacy en vrijheid van meningsuiting in beginsel hetzelfde gewicht.[37]

**5    Hoe nu verder?**

Er is dus een probleem met bijzondere persoonsgegevens. Hoe nu verder? Ik verken vijf mogelijke oplossingen, maar iedere oplossing heeft nadelen. Eerst noem ik drie manieren waarop, als Google in hoger beroep gaat, de rechter zou kunnen proberen het bijzondere-gegevens-probleem te verkleinen.

Een eerste mogelijkheid is om Google een beroep te gunnen op de media-exceptie. In de *Satamedia*-zaak heeft het HvJ EU deze media-exceptie ruim geïnterpreteerd.[38] De media-exceptie van de Wbp luidt als volgt:

> Artikel 3
>
> 1 Deze wet is niet van toepassing op de verwerking van persoonsgegevens voor uitsluitend journalistieke, artistieke of literaire doeleinden, behoudens de overige bepalingen van dit hoofdstuk, alsmede de artikelen 6 tot en met 11, 13 tot en met 15, 25 en 49.

---

[37] EHRM 7 februari 2012, 39954/08 (Axel Springer AG/Duitsland); EHRM 7 februari 2012, 40660/08 en 60641/08 (Von Hannover/Duitsland), par. 100; EHRM 16 juli 2013, 33846/07 (Wêgrzynowski en Smolczewski/Polen), par. 56.

[38] HvJ EG 16 december 2008, C-73/07 (Tietosuojavaltuutettu/Satakunnan Markkinapörssi Oy en Satamedia Oy). Zie ook: EHRM 21 juli 2015, 931/13, Satamedia/Finland, met noot W. Hins, Mediaforum 2016-1, 1. Zie over de media-exceptie: W. Hins, 'De journalistieke exceptie en de bescherming van persoonsgegevens: laveren tussen twee grondrechten', *Mediaforum* 2013-4, p. 98-104; J.V.J. Van Hoboken, 'The Proposed Right to be Forgotten Seen from the Perspective of Our Right to Remember, Freedom of Expression Safeguards in a Converging Information Environment', juni 2013, www.law.nyu.edu/sites/default/files/upload_documents/VanHoboken_RightTo%20Be%20Forgotten_Manuscript_2013.pdf.

> 2 Het verbod om persoonsgegevens als bedoeld in artikel 16 te verwerken is niet van toepassing voor zover dit noodzakelijk is voor de doeleinden als bedoeld in het eerste lid.

Een probleem met de media-exceptie-oplossing is dat het HvJ EU heeft gezegd dat Google zich daar niet op kan beroepen. In de authentieke versie van het Google Spain-arrest, in het Spaans, zegt het HvJ EU dat Google zich niet op de media-exceptie kan beroepen.[39] In de Nederlandse versie staat hetzelfde. De Engelse versie van het arrest laat de deur op een kier staan: het HvJ EU zegt dat het er niet op lijkt ('does not appear') dat Google zich op de media-exceptie kan beroepen. Maar dat lijkt een vertaalfout.[40]

Een ander probleem met de media-exceptie-oplossing is dat artikel 3 Wbp een beroep op het verwijderingsrecht helemaal zou uitsluiten.[41] Dat zou te ver gaan. Soms zouden mensen wel een recht moeten hebben om Google een link te laten verwijderen bij een zoekopdracht op naam; denk bijvoorbeeld aan wraakporno.[42]

Een tweede mogelijkheid om het bijzondere-gegevens-probleem op te lossen is een evenwicht vinden tussen de relevante grondrechten. Het toepassen van het persoonsgegevensbeschermingsrecht zou nooit een te grote inbreuk mogen maken op (andere) grondrechten, zoals het recht op vrijheid van meningsuiting.[43]

---

[39] HvJ EU 13 mei 2014, C-131/12, ECLI:EU:C:2014:317 (Google Spain), par. 85. Zie Art. 41 van de Rules of Procedure of the Court of Justice of the European Union: de Spaanse versie van het vonnis is authentiek. In de Spaanse versie van het arrest zegt het HvJ EU dat Google zich niet op de media-exceptie kan beroepen: '*ése no es el caso en el supuesto del tratamiento que lleva a cabo el gestor de un motor de búsqueda*'

[40] Zie over het arrest in verschillende talen ook: S. Kulk & F.J. Zuiderveen Borgesius, 'Freedom of Expression and "Right to Be Forgotten" Cases in the Netherlands after Google Spain', *European Data Protection Law Review* 2015-2, p. 113-125.

[41] Bij toepassing van artikel 3 Wbp hebben betrokkenen geen recht op verwijdering (artikel 36) en verzet (artikel 40). Ook sluit artikel 3 Wbp een verzoek tot bemiddeling door de Autoriteit Persoonsgegevens uit (artikel 47 Wbp). Dit probleem is veroorzaakt door de wijze waarop Nederland de Richtlijn heeft geïmplementeerd in de Wbp.

[42] Zie over wraakporno (revenge porn): D. Citron, *Hate crimes in cyberspace*, Cambridge: Harvard University Press 2014.

[43] In het Lindqvist-arrest zegt het HvJ EU dat nationale autoriteiten en rechters het gegevensbeschermingsrecht zo moeten toepassen dat een juist evenwicht verzekerd wordt tussen de verschillende grondrechten. HvJ EU 6 november 2003, ECLI:EU:C:2003:596 (Lindqvist), par. 90 en dictum.

Onder bepaalde voorwaarden mogen de rechten uit het EU Handvest, en dus het recht op bescherming van persoonsgegevens (artikel 8), beperkt worden. Het EU Handvest bevat een beperkingsclausule. Artikel 52 van het Handvest leest: 'Met inachtneming van het evenredigheidsbeginsel kunnen alleen beperkingen worden gesteld indien zij noodzakelijk zijn en daadwerkelijk aan door de Unie erkende doelstellingen van algemeen belang of aan de eisen van de bescherming van de rechten en vrijheden van anderen beantwoorden.'[44]

Een rechter zou kunnen beslissen dat de Wbp soms buiten toepassing moet blijven om rechten van anderen (zoals vrijheid van meningsuiting) te beschermen. Dat zou een soort externe toetsing, of een beperking van buitenaf, van de Wbp opleveren. Op die manier kan de vrijheid van meningsuiting beschermd worden bij verwijderingsverzoeken met betrekking tot bijzondere persoonsgegevens.

Maar het lijkt vreemd om de Wbp als geheel in te perken, terwijl de Wbp in theorie zelf al ruimte zou moeten bieden om de vrijheid van meningsuiting mee te wegen.[45] Bij niet-bijzondere persoonsgegevens kan informatievrijheid immers worden meegewogen met behulp van de legitiem-belang-bepaling en de media-exceptie.[46]

Een derde mogelijke oplossing voor het bijzondere-gegevens-probleem is een relatieve benadering van 'bijzondere persoonsgegevens'.[47] Een rechter zou kunnen betogen dat de persoonsgegevens voor de originele auteur (in dit geval de blogger) wél bijzondere gegevens zijn, maar dat dezelfde persoonsgegevens in handen van Google geen bijzondere persoonsgegevens zijn. De rechter zou dan kunnen redeneren dat Google niet de bedoeling heeft om bijzondere persoonsgegevens te verwerken. Het is echter onlogisch om te zeggen dat gegevens voor de ene verantwoordelijke wél bijzonder zijn, en voor de ander niet. Overigens lijkt uit jurisprudentie van de Hoge Raad te volgen dat de bedoeling van de verantwoordelijke geen rol speelt bij de beoordeling of persoonsgegevens bijzonder zijn.[48]

---

[44] Zie uitgebreid over de relatie tussen de beperkingsclausule van het Handvest (artikel 52) en het recht op bescherming van persoonsgegevens (artikel 8): G. González Fuster G & S. Gutwirth, 'Opening up Personal Data Protection: A Conceptual Controversy' (2013) 29(5) Computer Law & Security Review 531.
[45] Volgens het HvJ EU kan de vrijheid van meningsuiting binnen het gegevensbeschermingsrecht gerespecteerd worden: HvJ EU 6 november 2003, ECLI:EU:C:2003:596 (Lindqvist), par. 90 en dictum.
[46] Zie voor een vergelijkbare discussie omtrent externe toetsing van het auteursrecht: P.B. Hugenholtz, 'Auteursrecht contra informatievrijheid in Europa', in: A.W. Hins & A.J. Nieuwenhuis (red.), *Van ontvanger naar zender,* Amsterdam: Otto Cramwinckel 2003, p. 157-174; E.J. Dommering, 'De zaak Scarlet/Sabam: Naar een horizontale integratie van het auteursrecht', *AMI*, 2012, p. 49-53.
[47] Zie over het verschil tussen een relatieve en absolute benadering: F.J. Zuiderveen Borgesius, 'Mensen aanwijzen maar niet bij naam noemen: behavioural targeting, persoonsgegevens en de nieuwe Privacyverordening', *Tijdschrift voor Consumentenrecht en handelspraktijken* 2016-2, p. 54-66.

Afgezien van hoger beroep, zijn er nog twee mogelijkheden om het bijzondere-persoons-gegevens-probleem te verkleinen. Ten eerste kan de Autoriteit Persoonsgegevens een ontheffing verlenen voor het verwerken van bijzondere persoonsgegevens. De Wbp zegt dat het verbod op het verwerken van bijzondere persoonsgegevens niet van toepassing is voor zover 'dit noodzakelijk is met het oog op een zwaarwegend algemeen belang, passende waarborgen worden geboden ter bescherming van de persoonlijke levenssfeer en dit bij wet wordt bepaald dan wel het College [de Autoriteit Persoonsgegevens] ontheffing heeft verleend. Het College kan bij de verlening van ontheffing beperkingen en voorschriften opleggen'.[49] Zwenne heeft gesuggereerd dat Google zo'n ontheffing zou kunnen vragen aan de Autoriteit Persoonsgegevens.[50] Deze uitweg hangt dus af van het initiatief van Google en de reactie van de Autoriteit Persoonsgegevens.

Nóg een mogelijkheid is dat de wetgever ingrijpt. De Nederlandse wetgever zou de Wbp zo kunnen aanpassen dat er meer ruimte is voor vrijheid van meningsuiting. De wetgever moet de Wbp sowieso drastisch aanpassen, omdat in 2018 de Algemene Verordening Gegevensbescherming in werking treedt.[51] De Verordening verplicht de lidstaten het recht op bescherming van persoonsgegevens in overeenstemming te brengen met het recht op vrijheid van meningsuiting.[52] Dat wordt een lastige klus voor de lidstaten.

---

[48] HR 23 maart 2010, ECLI:NL:HR:2010:BK6331, r.o. 2.6. Zie: G-J. Zwenne & L. Mommers, 'Zijn foto's en beeldopnamen "rasgegevens" in de zin van artikel 126nd Sv en artikel 18 Wbp?', *Privacy & Informatie* 2010/5, p. 237247.
[49] Artikel 23(1)(f) Wbp.
[50] G.J. Zwenne, 'Het internetvergeetrecht', *Ars Aequi* 2015, afl. 1, p. 19.
[51] Verordening (EU) 2016/679 van het Europees Parlement en de Raad van 27 april 2016 betreffende de bescherming van natuurlijke personen in verband met de verwerking van persoonsgegevens en betreffende het vrije verkeer van die gegevens en tot intrekking van Richtlijn 95/46/EG (Algemene Verordening Gegevensbescherming) (*PbEG* L119/1).

Kortom, de Rechtbank Rotterdam heeft een beslissing genomen die in dit specifieke geval te begrijpen is. Maar de redenering van de rechtbank over bijzondere gegevens leidt tot problemen voor de vrijheid van meningsuiting. Het toepassen van de Wbp zou het recht op vrijheid van meningsuiting niet categorisch opzij mogen schuiven. Het is nu wachten op een creatieve rechter in hoger beroep.

∗ ∗ ∗

---

[52] Artikel 85 van de Algemene Verordening Gegevensbescherming.